\documentstyle[amssymb,prb,aps]{revtex}

\begin{document}
\title{Pinning of the domain walls of the cluster spin-glass phase in LTT
La$_{2-x}$Ba$_{x}$CuO$_{4}$}
\author{ F.Cordero,$^{1}$ A. Paolone,$^{2}$ R. Cantelli,$^{2}$ and M. 
Ferretti$^{3}$}
\address{$^{1}$CNR, Istituto di Acustica ``O.M. Corbino``, Via del Fosso del
Cavaliere 100,\\ I-00133 Roma, Italy and Unit\`{a} INFM-Roma1, P.le A. Moro 2,%
I-00185 Roma, Italy}
\address{$^{2}$Universit\`{a} di Roma ``La Sapienza``, Dipartimento di Fisica, 
and\\
Unit\`{a} INFM-Roma 1, P.le A. Moro 2, I-00185 Roma, Italy}
\address{$^{3}$Universit\`{a} di Genova, Dipartimento di Chimica e Chimica
Industriale,\\ and Unit\`{a} INFM-Genova, Via Dodecanneso 31, I-16146 Genova,
Italy}
\maketitle

\begin{abstract}
We compare the low frequency ($\sim 1$~kHz) anelastic spectra of La$_{2-x}$Sr%
$_{x}$CuO$_{4}$ and La$_{2-x} $Ba$_{x}$CuO$_{4}$ at $x=$ 0.03 and 0.06 in
the temperature region where the freezing into the cluster spin-glass (CSG)
phase occurs and is accompanied by an increase of the acoustic absorption.
The dependence of the amplitude of the anelastic relaxation on doping is
explained in terms of movement of the domain walls (DW) in the CSG phase
between the Sr (Ba) pinning points. The LBCO sample at $x=0.06$ transforms
into the LTT structure below 40~K and the amplitude of the anelastic anomaly
is 7 times smaller than expected, indicating pinning of the DW which run
parallel to the LTT modulation. Such DW can be identified with the stripes
of high hole density, and the present measurements show that they are mobile
between the Sr (Ba) pinning points down to few kelvin, but become static in
the presence of LTT modulation also far from the condition $x=\frac{1}{8}$
for commensurability between stripe and lattice periodicities.
\end{abstract}

\draft

The high temperature superconductors (HTSC) are a rich source of studies on
different aspects of condensed matter theories, including strong
electron-electron interactions and magnetism. Soon after the discovery of
HTSC, it was pointed out \cite{VSM87} that La$_{2}$CuO$_{4}$ was the first
experimental realization of a model $S=1/2$ quantum (Q) 2-dimensional (2D)
Heisenberg (H) Antiferromagnet (AF). HTSC are obtained from 2DQHAFs by
injecting holes into the Cu-O planes. The effect of doping in 2DQHAFs has
been widely studied (for a review see e.g. Ref. \onlinecite{RBC98}). In
particular it has been shown that in La$_{2-x}$M$_{x}$CuO$_{4}$, by
substituting M = Sr$^{2+}$ or Ba$^{2+}$ for La$^{3+}$, the long range AF
order is rapidly destroyed and $T_{N}$ decreases from 325$~$K to practically
0$~$K around $x_{c}\simeq 0.02$. When cooling below 30~K, the holes tend to
be localized over the Cu-O square plaquette near the Sr$^{2+}$ impurities,%
\cite{CBJ92,BCC95} allowing for the freezing of the spin degrees of freedom.
For $x<x_{c}$, a freezing of the in-plane spin distortions generated by
localized carriers is observed below $T_{f}\left( x\right) \simeq \left( 815~%
\text{K}\right) x$, corresponding to the onset of a spin-glass (SG) state,%
\cite{CBJ92,BCC95} while for $x>x_{c}$ the freezing into a cluster
spin-glass (CSG) phase is proposed below\cite{Joh97} $T_{g}\left( x\right)
=\left( 0.2~\text{K}\right) /x$. The SG state corresponds to local
distortions of the spin texture in AF background, while in the CSG phase
there are AF nanodomains separated by hole-rich domain walls (DW). The SG
and CSG states have been investigated by various experimental techniques,
such as NQR,\cite{RBC98,CBJ92,JBC99} $\mu $SR,\cite{NBB98} magnetic
susceptibility,\cite{CBK95} and recently anelastic spectroscopy.\cite{79}
Gooding {\it et al.}\cite{GSB97} have shown that, starting from a 2D
Heisenberg Hamiltonian and taking into account the role of dopant disorder,
at sufficiently low temperature the ground state of the system is a cluster
spin-glass. The effect of doping on the magnetic properties of HTSC, which
are good representatives of 2DQAF's, have been mainly restricted to the
study of La$_{2}$CuO$_{4}$ doped by Sr or Zn atoms. Only recently \cite
{NBB98} it has been shown that a common magnetic phase diagram as a function
of the hole concentration $n_{\text{sh}}$ in the CuO$_{2}$ sheets can be
drawn for La$_{2-x}$Sr$_{x}$CuO$_{4}$ and Y$_{1-x}$Ca$_{x}$Ba$_{2}$Cu$_{3}$O$%
_{6}$.

Closely connected with these phenomena is the formation of charged stripes,
which can be identified with the walls separating the hole-poor AF domains.%
\cite{JBC99,GSB97,TSA} In LSCO the ordering of the spins and consequently of
the charges into parallel stripes is dynamic and incommensurate with the
lattice, with a modulation wave vector shifted from the AF wave vector $q_{%
{\rm AF}}=\left( \frac{1}{2},\frac{1}{2}\right) $ by $\delta \simeq x=n_{%
\text{sh}}$ (in the tetragonal notation). Such a linear relationship between
incommensurability $\delta $ and doping is found to hold for $0.04<x<1/8$,
while $\delta $ remains locked at $\sim 1/8$ at higher doping.\cite{YLK98}
The magnetic modulation is parallel to the modulation of the CuO$_{2}$ plane
in the LTO structure (at 45$^{\text{o}}$ with the Cu-O bonds) for $x\le 0.05$%
, but becomes parallel to the Cu-O bonds across the metal-insulator
transition, $x\ge 0.06$ (Ref. \onlinecite{WBK00}). These magnetic neutron
diffraction peaks have been recently reported to be elastic,\cite{WBK00}
although they could actually be quasielastic and therefore correspond to
fluctuating correlations which are observable only by fast probes like
neutrons.\cite{WSE99} These correlations are considered to be static in
samples where the substitution of La with both Sr ($0.12\le x\le 0.20$) and
Nd stabilizes the low-temperature tetragonal (LTT) structure, with lattice
modulation parallel to the stripes;\cite{TSA,FGY00} at such doping levels
the stripe wave vector is also close to the commensurate value $\frac{1}{8}$.

In order to obtain further insight in this complex interplay of dynamic
magnetic and charge ordering and pinning by the lattice, we extended the
acoustic spectroscopy measurements of the CSG state in LSCO (Ref. %
\onlinecite{79}) to La$_{2-x}$Ba$_{{x}}$CuO$_{4}$ (LBCO), which has greater
tendency to form a stable LTT phase. The present study demonstrates that the
domain walls of the CSG phase (which can be identified as the hole stripes)\
become static within the LTT structure of LBCO even for doping at which the
stripe spacing is not commensurate with the lattice.

Two La$_{2-x}$Ba$_{{x}}$CuO$_{4}$\ samples and two La$_{2-x}$Sr$_{x}$CuO$%
_{4} $\ samples with $x=0.03$\ and $x=0.06$\ were prepared by standard solid
state reaction as described in Ref.~\onlinecite{NGM99} and cut in bars
approximately $45\times 5\times 0.6$\ mm$^{3}$. The complex Young's modulus $%
E\left( \omega \right) =E^{\prime }+iE^{\prime \prime }$, whose reciprocal
is the elastic compliance $s=E^{-1}$, was measured as a function of
temperature by electrostatically exciting the lowest three odd flexural
modes and detecting the vibration amplitude by a frequency modulation
technique. The vibration frequency, $\omega /2\pi $, is proportional to $%
\sqrt{E^{\prime }}$, while the elastic energy loss coefficient (or
reciprocal of the mechanical $Q$) is given by\cite{NB} $Q^{-1}\left( \omega
,T\right) =$ $E^{\prime \prime }/E^{\prime }=$ $s^{\prime \prime }/s^{\prime
}$, and was measured by the decay of the free oscillations or the width of
the resonance peak. The imaginary part of the dynamic susceptibility $%
s^{\prime \prime }$ is related to the spectral density $J_{\varepsilon
}\left( \omega ,T\right) =$ $\int dt\,e^{i\omega t}\left\langle \varepsilon
\left( t\right) \varepsilon \left( 0\right) \right\rangle $ of the
macroscopic strain $\varepsilon $ through the fluctuation-dissipation
theorem, $s^{\prime \prime }=$ $\left( \omega V/2k_{\text{B}}T\right)
J_{\varepsilon }$.

The spin dynamics can affect $J_{\varepsilon }$ through magnetoelastic
coupling,\cite{NB} namely an anisotropic strain with principal axes oriented
according to the spin orientation. Indeed, it has recently been shown\cite
{79} that in LSCO the freezing into the CSG state is accompanied by a
step-like increase of the acoustic absorption and a rise of the elastic
modulus; the data of La$_{2-x}$Sr$_{x}$CuO$_{4}$ with $x=0.03$ and 0.06 are
shown in Figs. 1 and 2. The anomaly is observed around or slightly below $%
T_{g}$, as reported in the literature\cite{Joh97} ($T_{g}\left(
x=0.03\right) \simeq 6.7$~K and $T_{g}\left( x=0.06\right) \simeq 3.3$~K)
and deduced from the maxima in the $^{139}$La NQR relaxation rate measured
on the same samples.\cite{79,82} The NQR and $\mu $SR experiments measure
the spectral density $J_{\text{spin}}\left( \omega ,T\right) $ of the spin
fluctuations. If the fluctuations occur with a rate $\tau ^{-1}\left(
T\right) $, then $J_{\text{spin}}\propto \tau /\left[ 1+\left( \omega \tau
\right) ^{2}\right] $ is peaked at the temperature at which $\omega \tau =1,$
and the sharp peaks in the NQR and $\mu $SR relaxation rates indicate the
slowing of the spin fluctuations below the measuring angular frequency $%
\omega $\ ($\sim 10^{7}-$ $10^{8}$~Hz in those experiments). The same peak,
shifted to lower temperature because of the lower frequency ($\omega \sim $ $%
10^{3}-$ $10^{4}$~Hz), could in principle appear also in $J_{\varepsilon }$
thanks to the magnetoelastic coupling. However, the step-like $Q^{-1}\left(
T\right) $ curves\ of Fig. 1 cannot be easily identified with the sharp
peaks in the NQR\ rate,\cite{RBC98,CBJ92,JBC99,82} and can be attributed to
the movement of domain walls separating the antiferromagnetically correlated
domains.\cite{79} In this case, the steep rise of the absorption with
lowering temperature is determined by the formation of the clusters with
frozen AF spin correlations, while the broader shape at lower temperature is
due to the wide distribution of the relaxation rates of domain walls with
different lengths and mobilities.

To our knowledge, the low-doping and low- temperature region of the magnetic
phase diagram of \thinspace LBCO has not been investigated, but the close
similarity between the anelastic spectra of LBCO and LSCO in Figs. 1 and 2
indicate that also in LBCO\ the spins undergo the same freezing processes at
practically the same temperatures as in LSCO. The anomalies in the imaginary
(Fig. 1)\ and real parts (Fig. 2) of the dynamic elastic response attributed
to the appearance of the CSG state appear below 7~K for $x=0.03$ in both
samples and below 4~K (3~K) for $x=0.06$ in LSCO (LBCO). The similarity
between the spectra at $x=0.03$ is shown in Fig. 3, where both the
absorption and real parts of the dynamic response of LSCO and LBCO are
plotted together (in all figures the Young's modulus $E\left( T\right) $ is
normalized to the extrapolated $E\left( 0\right) $). The data of LSCO
practically overlap with those of LBCO if multiplied by 1.45, indicating
that the rise of the elastic modulus is determined by the same mechanism
producing the absorption. The factor $\sim 1.5$ can be ascribed to a greater
magnetoelastic coupling in the Ba compound, possibly resulting from the
slightly different atomic sizes and distances.

We turn now to the dependence of the amplitude of the anelastic effect on
doping. For LSCO, the amplitude is a decreasing function of doping, as shown
by the open circles of Fig. 4, which includes also data from Ref. %
\onlinecite{79}. We want to show that this trend is in agreement with the
proposed mechanism of dissipation (and consequent dispersion in the real
part of the modulus\cite{NB}) due to the stress-induced change of the sizes of
domains with differently oriented axes of staggered magnetization.\cite{79}
Since the Sr dopants act as pinning points for the DW, their concentration
affects not only the density $n$ of DW but also their mean length $l$
between pinning points. The relaxation strength $\Delta $ is expected to be
of the form $\Delta \propto n\,l^{\alpha }$ with $\alpha =3$, similarly to
the case of the motion of dislocations pinned by impurities,\cite{NB} or
with $\alpha =2$, as proposed for the susceptibility due to the motion of DW
between ferromagnetic domains.\cite{NSV90}

In Fig. 4 the relaxation strength measured with LSCO is compared with $%
\Delta \left( x\right) \propto n\left( x\right) \,l^{\alpha }\left( x\right) 
$ calculated assuming that the Sr atoms above and below a CuO$_{2}$ plane
form a square lattice with spacing $d=1/\sqrt{x}$ in units of the lattice
constant $a$, and that the DW form a square lattice passing through the Sr
atoms; the mean number of free atoms along a DW segment is then $l=d-1$,
since the two extrema at the Sr pinning points are fixed. In this manner we
obtain $\Delta \propto 2x\left( \frac{1}{\sqrt{x}}-1\right) ^{\alpha }$,
which reproduces reasonably well the experimental data of LSCO with $\alpha
\lesssim 3$. If we assume that the DW run only parallel to one direction, as
expected from parallel stripes, then $n\left( x\right) $ is halved, but the
dependence on $x$ remains the same. We conclude that the dependence of the
relaxation magnitude on doping is in agreement with what expected from the
motion of DW pinned by the Sr atoms.

The case of LBCO at $x=0.06$ is different, since the amplitude of the step
at $T_{g}\sim 3.5$~K is 4.8 times smaller than for LSCO (both in the
absorption and real parts), and the onset temperature is also slightly lower
(Fig. 1). Considering the difference in magnetoelastic coupling, it turns
out that the amplitude of the effect in LBCO at $x=0.06$ is 7 times smaller
than expected. This reduction appears clearly in Fig. 4, where all the data
are normalized at $x=0.03$. The depression of the absorption step in LBCO
with $x=0.06$ has to be attributed to the transformation into the LTT
structure below $T_{t}\simeq 40$~K, which is observed in the anelastic
spectrum (Figs. 1 and 2). The LTO-LTT transition has been mainly studied on
samples with $x\simeq 0.12$,\cite{AMH89} in view of the strong decrease of $%
T_{c}$\ around that doping. To our$~$knowledge no data are available for $%
0.05\leq x\leq 0.07$ in LBCO, but our observation of $T_{t}\simeq 40$~K at $%
x=0.06$ is in agreement with the available data\cite{AMH89,PR91} of $T_{t}=0$%
~K for $x\leq 0.05$\ and $T_{t}=50$~K for $x=0.08$. The effect of the
LTO-LTT transition on the acoustic properties of LBCO has been identified%
\cite{FNH90} in a small sound attenuation peak and lattice stiffening below $%
T_{t}$\ on LBCO samples with $0.10\leq x\leq 0.16$. The amplitude of the
stiffening,$\ \Delta E/E$,\ below $T_{t}$\ is strongly doping dependent,
with a maximum of 0.05 around $x=0.12$,\cite{FNH90} and already lowered to
0.01 at $x=0.10$; the value of 0.004 which we find at $x=0.06$\ is
consistent with this trend.{\em \ }Therefore, we identify the increase of $%
E^{\prime }$\ below $T_{t}\sim 40$~K with the onset of the transformation to
the LTT structure.

The fact that the anelastic relaxation due to the DW motion in the LTT
structure of LBCO is 7 times smaller than in the LTO one of LSCO is a clear
indication of pinning of the DW motion within the LTT phase. The pinning of
the DW can be attributed to the LTT lattice modulation, with rows of
inequivalent O atoms in the CuO$_{2}$ plane (within or out of the plane)
along the direction of the Cu-O bonds, irrespective of commensuration
effects. Only the walls parallel or nearly parallel to this modulation
should be affected, and therefore from the reduction factor $\sim 7$ we
deduce that about 87\% of the DW are parallel to the direction of the Cu-O
bonds in the LTT phase (in the assumption of a complete transition to the
LTT structure), in agreement with the direction of the magnetic
modulation observed by neutron scattering for $x\ge 0.055$.\cite{WBK00}

The pinning of the DW of the CSG phase within the LTT structure is not
unexpected, since the pinning of the dynamic charge and spin fluctuations
into static stripe modulations is well documented for samples with Ba or
mixed Nd and Sr substitution with $x\ge 0.12$, with modulation wave vector
close to the commensurate value $\frac{1}{8}$ and parallel to the LTT
lattice modulation.\cite{TSA} In the present case, however, $x=0.06$ is far
from the condition of commensurability between stripes and lattice. Also, we
emphasize that the suppression of the acoustic absorption from the DW motion
implies that the pinned DW are static over a time scale longer than $\omega
^{-1}\sim 10^{-3}$~s, which is a much more stringent condition than that for
the observation of elastic/quasielastic neutron diffraction peaks.\cite
{WSE99}

In conclusion, we compared the anelastic spectra of Sr- and Ba-substituted
La cuprate in the temperature region where the freezing into the cluster
spin-glass phase is observed, at doping 0.03 and 0.06. The dependence of the
magnitude of the anelastic anomaly on the concentration of Sr has been
quantitatively explained in terms of motion of DW which are pinned by the Sr
substitutional atoms. The anelastic spectra at $x=0.03$ , with both LSCO and
LBCO having LTO\ structure, completely overlap with a scaling factor of
1.45, attributed to greater magnetoelastic coupling strength with Ba
substitution. Instead, at $x=0.06$ LBCO transforms into the LTT structure
and the acoustic absorption due to the motion of the AF domain walls in the
CSG phase is strongly reduced, indicating nearly complete pinning of the
magnetic domain walls within the LTT structure.
The DW between different AF spin clusters are identified with stripes of
high hole density, and the rows of inequivalent O atoms in the LTT\
structure would provide a modulated potential which pins the charge stripes
parallel to this modulation. The present measurements show that this clamping
effect is strong also in the absence of commensuration between stripe and
lattice periodicities.

We thank R.S. Markiewicz for useful discussions. This work has been done in
the framework of the Advanced Research Project SPIS of INFM.


\section{Captions to figures}

Fig. 1 Elastic energy loss coefficient of LSCO and LBCO at doping $x=0.03$
and 0.06, measured at $\sim 1.5$~kHz.

Fig. 2 Relative variation of the Young's modulus of the samples of Fig. 1.
The rise of the modulus below 40~K in LBCO $x=0.06$ is attributed to the
formation of LTT phase.

Fig. 3 Comparison between the absorption and real part of the dynamic
Young's modulus of LSCO (1.7~kHz) and LBCO (1.2~kHz) at $x=0.03$ in the
region of the freezing into the CSG state. The curves of LSCO have been
multiplied by 1.45.

Fig. 4 Intensity $\Delta $ of the elastic energy loss contribution of the
domain walls in the CSG state as a function of doping for LSCO and LBCO;
the datum at $x=0.025$ is from Ref. \onlinecite{79}.
The continuous lines are
obtained from the expression of $\Delta \left( x\right) $ discussed in the
text. The data and the curves are normalized at $x=0.03$.

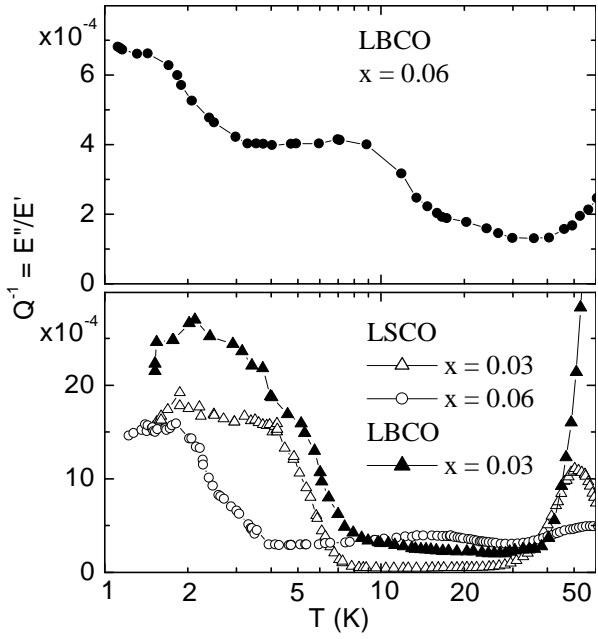
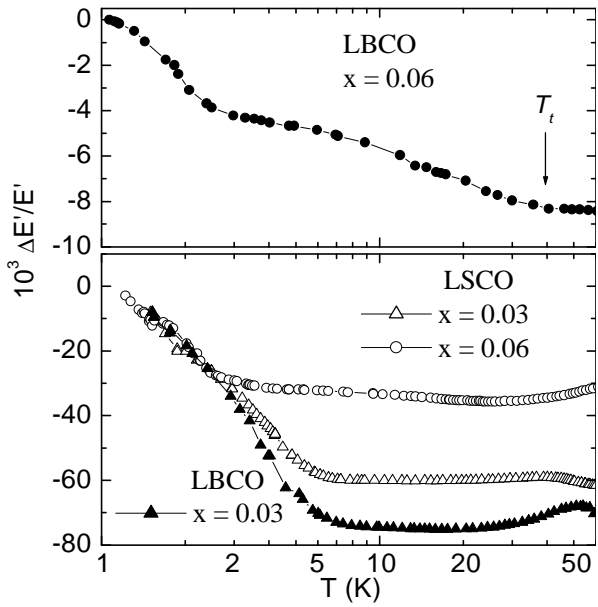
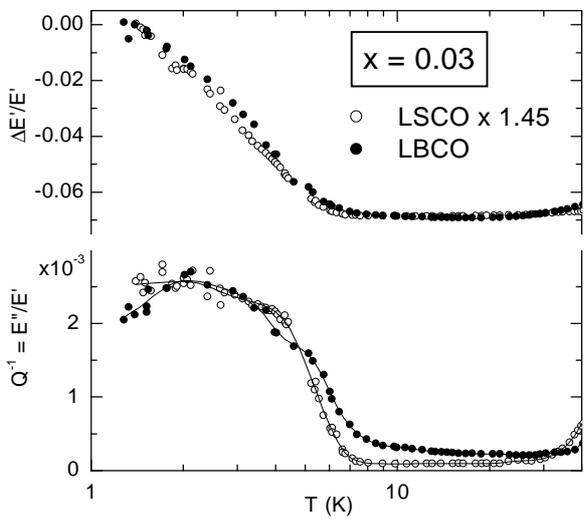
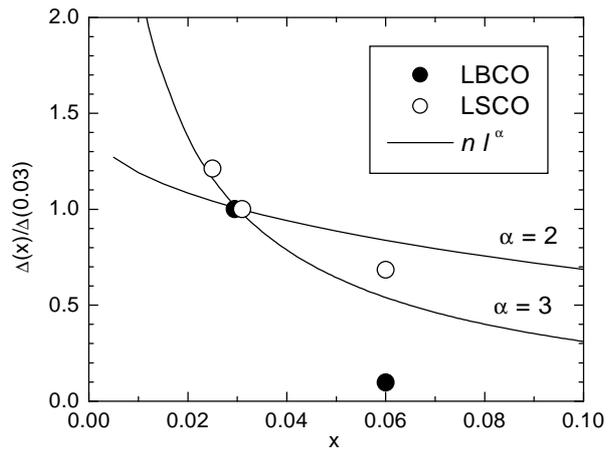
\end{document}